\input{aipcheck}

\documentclass[
    ,final            
  ]
  {aipproc}

\layoutstyle{6x9}

\begin{document}

\title{Probing Relativistic Astrophysics Around SMBHs: The Suzaku AGN Spin Survey}

\classification{98.54.Cm, 95.30.Sf, 95.85.Nv}
\keywords      {Active Galactic Nuclei, Seyfert galaxies, accretion disks, black holes}

\author{C.S.Reynolds}{
  address={Dept. of Astronomy, University of Maryland, College Park, MD20742, USA}
}

\author{L.W.Brenneman}{
  address={Harvard Smithsonian Center for Astrophysics, 60 Garden Street, Cambridge, MA, USA}
}

\author{A.M.Lohfink}{
  address={Dept. of Astronomy, University of Maryland, College Park, MD20742, USA}
}

\author{M.L.Trippe}{
  address={Dept. of Astronomy, University of Maryland, College Park, MD20742, USA}
}

\author{J.M.Miller}{
  address={Department of Astronomy, University of Michigan, Ann Arbor, MI 48109, USA}
}

\author{R.C.Reis}{
  address={Department of Astronomy, University of Michigan, Ann Arbor, MI 48109, USA}
}

\author{M.A.Nowak}{
  address={MIT Kavli Institute for Astrophysics, Cambridge, MA 02139, USA}
}

\author{A.C.Fabian}{
  address={Institute of Astronomy, Madingley Road, Cambridge, CB3 OHA, UK}
}

\begin{abstract}
In addition to providing vital clues as to the formation and evolution of black holes, the spin of black holes may be an important energy source in the Universe.  Over the past couple of years, tremendous progress has been made in the realm of observational measurements of spin.  In particular, detailed characterization and modeling of X-ray spectral features emitted from the inner regions of black hole accretion disks have allowed us to probe the location of the innermost stable circular orbit and hence the spin.   In this contribution, I will describe the {\it Suzaku AGN Spin Survey}, an on-going Cycle 4--6 {\it Suzaku} Key Project that aims to probe five nearby AGN with sufficient depth that strong gravity effects and spin can be robustly assessed.    Of the three objects for which we currently have our deep datasets, we can constrain spin in two (NGC~3783, $a>0.9$; Fairall~9, $a=0.45\pm 0.15$) whereas complexities with the warm absorber prevent us from reporting results for NGC~3516.  We conclude with a brief discussion of spin-dependent selection biases in flux-limited surveys.
\end{abstract}

\maketitle


\section{Introduction}

Probing the physics of the strong gravity region close to astrophysical black holes has been a long held promise of X-ray astronomy.  Starting with {\it ASCA} studies of the Seyfert galaxy MCG--6-30-15 \citep{tanaka95,fabian95}, studies of relativistically broadened and redshifted X-ray spectral features (aka broad iron lines) from the innermost regions of black hole accretion disks have allowed this promise to be realized.  In this contribution, I shall give a very brief introduction to broad iron line studies of active galactic nuclei (AGN) before giving a progress report on the {\it Suzaku AGN Spin Survey}, a Cycle 4--6 {\it Suzaku} Key Project whose primary aim is to measure the black hole spin in a sample of nearby AGN.   I shall conclude with a brief discussion of spin-dependent selection biases that may affect future flux-limited surveys of black hole spin.

\begin{figure}
  \includegraphics[width=.45\textwidth,angle=270]{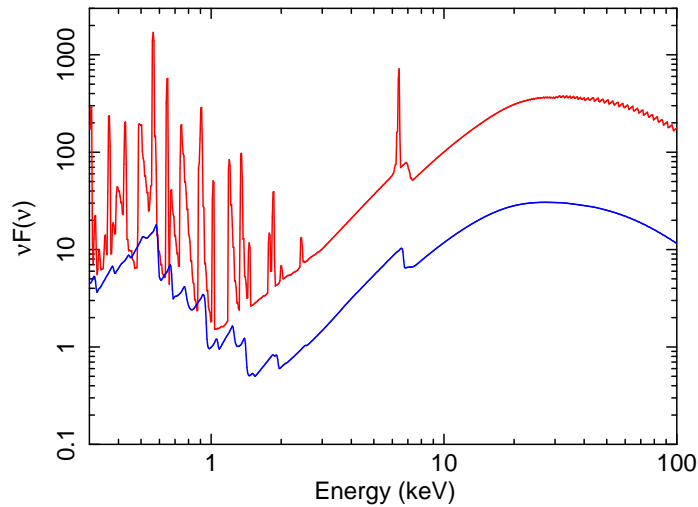}
  \caption{Theoretical X-ray reflection spectra (tabulated in the {\tt reflionx} model; \cite{ross05}).  Shown in red (top) is the rest-frame X-ray reflection spectrum assuming that the disk has an ionization parameter $\xi=10\,{\rm erg}\,{\rm s}^{-1}\,{\rm cm}$, solar abundances and is irradiated by a $\Gamma=2$ power-law spectrum of X-rays.   The blue (bottom) curve shows this same spectrum convolved with relativistic smearing relevant for a disk extending down to the ISCO of rapidly spinning ($a=0.9$) black hole viewed at 30\,degrees inclination with an irradiation profile of $\epsilon\propto r^{-3}$. For clarity, the two spectra have been offset in normalization by a factor of 10.}
  \label{fig:reflection}
\end{figure}

What are the origin of the spectral features that we study?    There is strong evidence that luminous AGN accrete from radiatively-efficient disks \citep{davis11,soltan82} in which the vast majority of the accretion energy is released within $r\sim 20GM/c^2$ of the supermassive black hole (SMBH).  Much of this energy is radiated as thermal optical/UV emission.  However, of course, AGN are known to be powerful sources of hard (variable) X-ray emission and so we conclude that a significant fraction of the accretion energy is channeled into an X-ray emitting corona situated above the cold, optically-thick disk.   Given this geometry, the inner accretion disk will be strongly irradiated by the X-ray corona resulting in the so-called X-ray reflection spectrum consisting of fluorescence lines, radiative recombination emission, and Compton backscattered radiation (Fig.~\ref{fig:reflection}).    The presence of these X-ray reflection features is a natural consequence of the basic AGN paradigm.   As expected, we see very similar features in the spectra of Galactic Black Hole Binaries \cite{miller09} and accreting weakly-magnetized neutron stars \cite{cackett10}.

Characterizing the X-ray reflection spectrum provides a our best window to date on the immediate environment of the SMBH.   Fitting full reflection models convolved with appropriate relativistic smearing kernels (Fig.~\ref{fig:reflection}) to high-quality broad-band X-ray spectra allow us to determine the photospheric ionization state, the photospheric iron abundance, the inclination, and the irradiation profile of the inner disk.   We can also determine the inner ``reflection" edge of the disk \citep{krolik02} from the strength of the gravitational redshifting.  Identifying this edge with the general relativistic innermost stable orbit then allows us to constrain the spin of the SMBH \citep{reynolds08}.

\section{The Suzaku AGN Spin Survey}

\subsection{Motivation and basic goals}

Extracting the relatively subtle signatures of black hole spin from X-ray spectra requires high-quality data.   For this reason, prior to the initiation of {\it Suzaku} Key Projects in 2008, only one AGN possessed a useful constraint on the spin of its SMBH; multiple analyzes of MCG--6-30-15 found a rapidly rotating black hole, with dimensionless spin parameter $a>0.9$ \citep{dabrowski97,brenneman06,miniutti07}.    Even this single datapoint proves interesting; the discovery of such a rapidly rotating black hole in a radio-quiet AGN rules out simple spin-driven models for the radio-quiet/radio-loud dichotomy \citep{sikora07}.   Clearly, expanding a spin study to a sample of AGN is a high priority.

The {\it Suzaku AGN Spin Survey} is an on-going Cycle 4--6 {\it Suzaku} Key Project that, when complete, will provide high-quality spectral datasets for five nearby AGN that had previous indications for broad iron lines (NGC3516, NGC~3783, Fairall~9, 3C120, and Mrk~841). The primary goals of this project are to study the physics of the innermost accretion disk and the black hole spin by characterizing the relativistically broadened reflection features.  However, these datasets also allow studies of the circumnuclear environment (via the emission/absorption signatures of photoionized matter) as well as the physics of the X-ray continuum (via the study of spectral curvature and energy-dependent temporal lags) and hence will be a valuable legacy to AGN researchers.

In the rest of this contribution, we describe the current results from this program.

\subsection{NGC~3783}

\begin{figure}
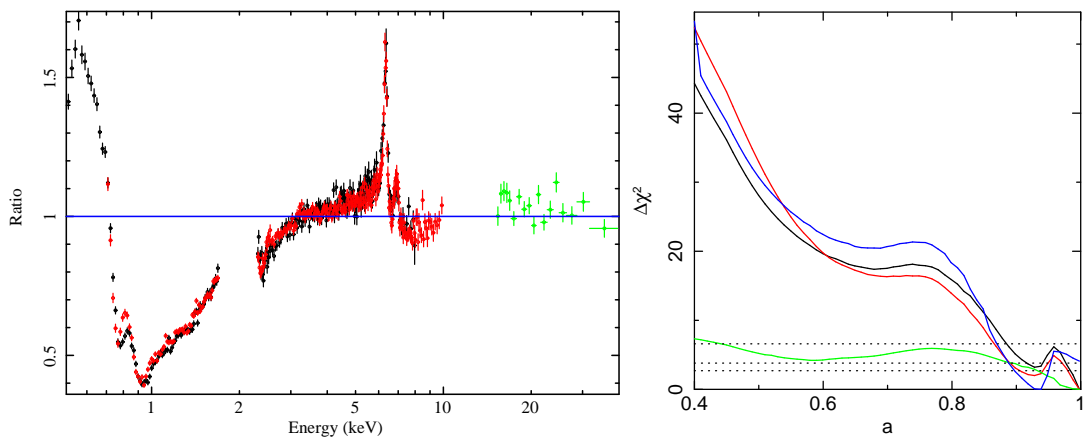

\hbox{
  \includegraphics[height=.37\textheight,angle=270]{ngc3783_ratio.ps}
  \includegraphics[height=.28\textheight,angle=270]{ngc3783_spin.eps}
  }
  \caption{{\it Left panel : }0.5--40\,keV {\it Suzaku}/XIS+PIN spectrum of NGC~3783 shown as a ratio against a simple powerlaw continuum.  XIS/BI data are shown in red, XIS/FI data are in black (co-added for the purposes of display), and PIN data are in green.  Note the strong warm absorber below 3\,keV and the strong narrow iron line at 6.4\,keV (rest frame).  Despite these spectral complexities, broadened disk reflection is strongly required by these data.  Figure from \cite{reis11}.  {\it Right panel : }Goodness of fit parameter as a function of dimensionless spin parameter for a fiducial spectral model which includes a multi-zone warm absorber, distant reflection, soft excess, disk reflection, and a fixed XIS/PIN cross-normalization factor (black line). Also shown are variants of this model where the warm absorber parameters are frozen (red; to assess trades with absorption parameters), the XIS/PIN cross-normalization is a free parameter (blue), and only hard band ($>3$\,keV) data are modeled (green).   From bottom to top, the dotted lines show the 90\%, 95\% and 99\% confidence levels for one interesting parameter. Figure from \cite{brenneman11}.  }
  \label{fig:ngc3783_spec}
\end{figure}

The Seyfert~1.5 galaxy NGC~3783 was observed quasi-continuously by {\it Suzaku} in July 2009, with a total on-source exposure of 210\,ks.  Detailed analysis of these data can be found in \cite{brenneman11,reis11}; here we give a brief summary of these findings.  Figure~\ref{fig:ngc3783_spec} (left) shows the 0.5--40\,keV {\it Suzaku}/XIS+PIN spectrum of this object.  Here, and in all {\it Suzaku} spectra presented here, we ignore XIS data in the 1.5--2.5\,keV range due to some obvious calibration problems associated with modeling the absorption edges in the X-ray mirror.  The soft X-ray spectrum ($<3$\,keV) is strongly affected by a warm absorber that has been well studied in previous works, including a 900\,ks {\it Chandra} High-Energy Transmission Grating (HETG) campaign in 2000--2001 \citep{netzer2003}.  A narrow fluorescent 6.4\,keV iron line from low velocity, distant material is also readily apparent.    Despite these  complexities, the spectrum strongly requires relativistically smeared reflection from the inner accretion disk \cite{brenneman11}.   Describing the disk reflection with the {\tt reflionx} ionized reflection model \cite{ross05}, and using the {\tt relconv} relativistic smearing model \cite{dauser10}, we find a disk inclination of $i=22^{+3}_{-8}$\,degrees and supersolar iron abundance $Z=3.7\pm 0.9Z_\odot$.  The relativistic broadening is very strong, requiring black hole spin of $a>0.9$ at the 90\% confidence level (CL) even allowing for possible systematic errors introduced by the analysis/modeling procedure (Fig.~2 [right]).

\begin{figure}
\hbox{
  \includegraphics[height=.4\textheight]{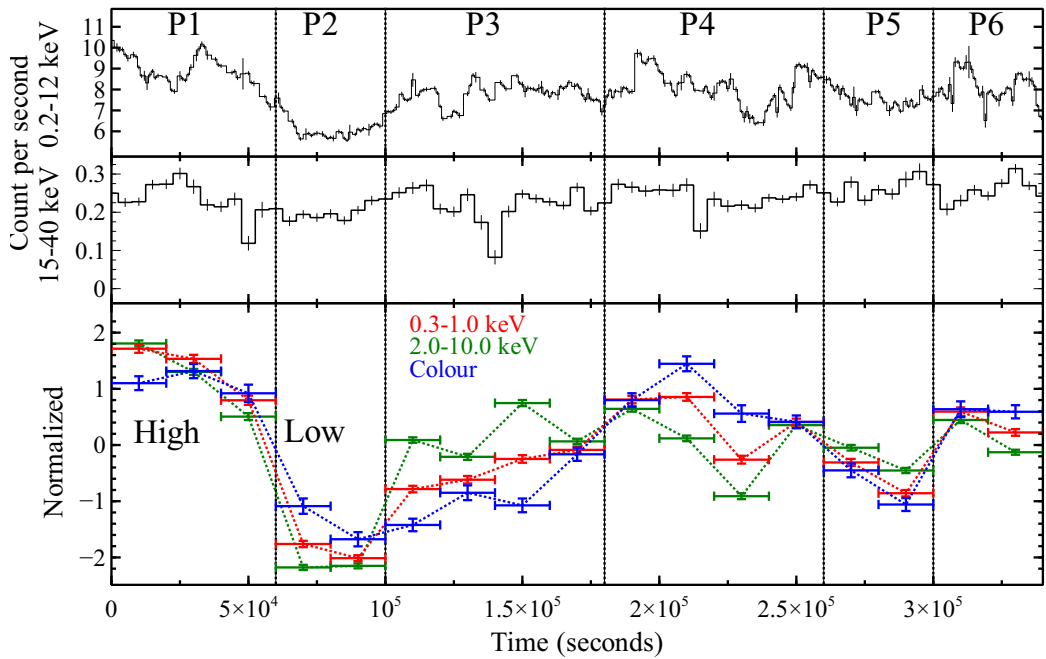}
}
  \caption{Summed XIS light curve for NGC~3783 (top panel) together with normalized restricted band light curves and the softness ratio (bottom panel).  Complex patterns of variability are revealed; during some periods (P1 and P5--P6), color variations track the count rate variations perfectly whereas during other periods (P3 and P4), significant color variations are seen which are uncorrelated with overall count rate.  Detailed analysis shows that the spectral variability can be explained as being due, principally, to changes in the ionization and strength of the inner disk reflection.  Figure from \cite{reis11}.}
  \label{fig:ngc3783_variability}
\end{figure}

We note briefly the independent analysis of these data by Patrick and collaborators \cite{patrick11} which, contrary to our conclusions, finds that NGC~3783 has a low (or even retrograde) spin, $a<0.37$.  This alternative analysis differs from ours in several key ways. Probably most importantly, they include an additional continuum component (described by a ``warm" comptonization model), and make the ansatz that the iron abundance is solar.  We have confirmed that, fixing the iron abundance to unity, significantly decreases the inferred spin (90\% confidence range $a=0.28-68$).  However, even when we include the additional comptonization component, we find that supersolar abundance is strongly required (90\% confidence range $Z=3.3-3.7\,Z_\odot$; $\Delta\chi^2=35$ for 1 extra degree of freedom) and the need for rapid prograde spin is restored.  Thus, we argue that the solar abundance ansatz can be falsified (we also note that the cores of Seyfert galaxies are atypical environments and hence supersolar abundances are not unreasonable \cite{orban11}).   A more complete comparison of our analysis with the Patrick analysis, including a discussion of some important differences in data handling and modeling, will be presented in a forthcoming publication.   

Significant flux and spectral variability is seen during this {\it Suzaku} observation (Fig.~\ref{fig:ngc3783_variability}); our detailed analysis of the variability is described in \cite{reis11}.  We find that the spectral variability (including large amplitude changes in the PIN-band flux) is driven by changes in the ionization and strength of the X-ray reflection from the inner accretion disk, whereas the warm absorber is remarkably constant during the 400\,ks duration of this observation.  Astrophysically, these changes must be a result of changes in both the power and geometry/location of the irradiating X-ray source.

\subsection{Fairall~9}

\begin{figure}
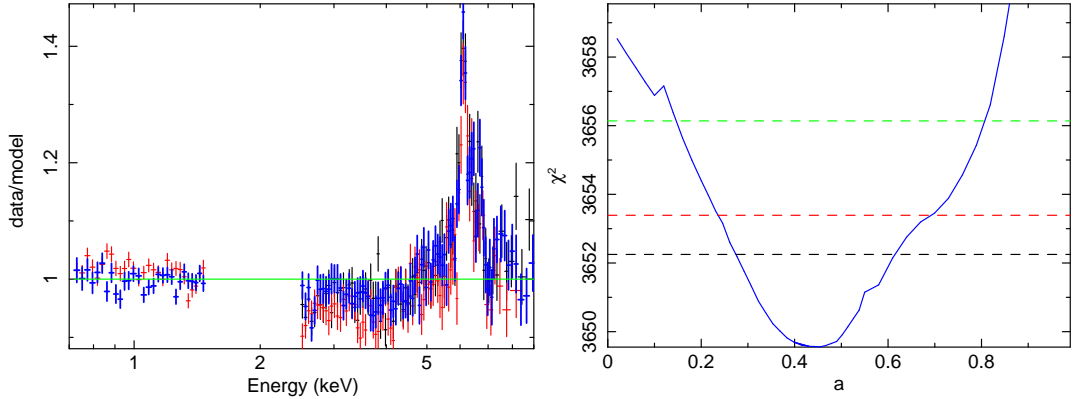

\hbox{
  \includegraphics[height=.32\textheight,angle=270]{f9_ratio.ps}
  \includegraphics[height=.32\textheight,angle=270]{f9_spin.ps}
  }
  \caption{{\it Left panel : }XIS0 (black), XIS1 (red) and XIS3 (blue) spectra of Fairall~9 ratioed against a simple unabsorbed power-law model.  Note the strong narrow and broad iron lines in the spectrum of this remarkable ``clean" (i.e. unabsorbed) AGN.  {\it Right panel : }Preliminary spin constraint on Fairall~9.  Shown here is the goodness of fit parameter as a function of black hole spin in a multi-epoch analysis of the Key Project {\it Suzaku} data along with data from the long 2010 {\it XMM-Newton} observation.  The spectral model includes FeXXV and FeXXVI line emission from photoionized plasma in the circumnuclear environment (assumed to remain constant between the {\it Suzaku} and {\it XMM-Newton} observations) and a soft excess component described by an additional thermal comptonization model.    See A.Lohfink's contribution in this volume for further details.}
  \label{fig:f9}
\end{figure}

The luminous Seyfert 1 galaxy Fairall~9 was observed quasi-continuously in May 2010, with a total on-source exposure of 190\,ks.  Fairall~9 is a prototypical example of a ``bare" Seyfert nucleus \cite{patrick10}; previous studies have failed to find any signatures of either neutral or ionized absorption.   Consequently, this is a particularly clean object in which to study relativistic physics.   Figure~\ref{fig:f9} (left) shows a simple ratio of the (three) XIS spectra to an unabsorbed power-law model.    Both strong narrow and broad iron lines are seen clearly.

A detailed discussion of this dataset is given in Anne Lohfink's contribution in this volume as well as in \cite{lohfink11}.   Here, we simply note that a detailed study of the inner disk and black hole spin requires us to model additional (narrow) FeXXV and FeXXVI line emission as well as a subtle soft excess component.   Degeneracies between these components exist when we model these {\it Suzaku} data in isolation, but these degeneracies are largely resolved if we perform a joint analysis of these data with the 2010 deep {\it XMM-Newton} pointing \cite{emm11} .  Work is still on-going and will be reported in a future publication, but our preliminary black hole spin of $a=0.45\pm 0.15$ (90\% CL; see Fig.~\ref{fig:f9}) is completely consistent with previous measurements of spin in this object by Schmoll et al.\cite{schmoll09}

\subsection{NGC~3516}

\begin{figure}
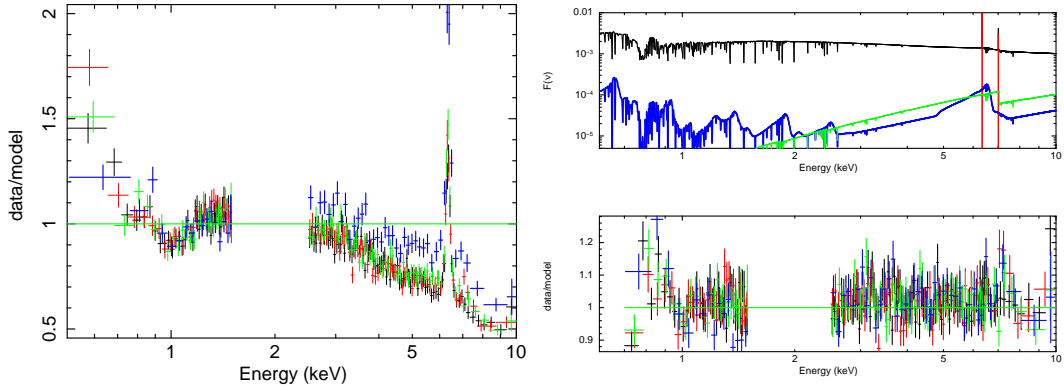

\hbox{
  \includegraphics[height=.32\textheight,angle=270]{ngc3516_tslice_ratio_pl.ps}
  \includegraphics[height=.32\textheight,angle=270]{ngc3516_broadline.ps}
  }
  \caption{XIS spectra of NGC~3516.   {\it Left panel : }Ratio of XIS0 spectra to a simple unabsorbed power-law model for four consecutive segments of data (black$\rightarrow$red$\rightarrow$green$\rightarrow$blue).  The overall flux undergoes an almost monotonic decline by a factor of three during this time.    Each segment has approximately 60\,ks of on-source exposure, with a total wall-clock duration of 90\,ks.  The photon index of the reference power-law is tied between all four fits.  The warm absorber imprints a clear ``notch" at 1\,keV, and there is no evidence for any change in this absorber between the four segments.   A strong narrow iron line is seen in all segments.  The spectral hardening of the final (blue; faintest) data segment can be completely attributed to the reflection continuum of a constant (distant) reflector.  {\it Right panel : }Spectral fits to these four data segments of a model consisting of a powerlaw with a 2-zone warm absorber plus distant reflection (green/red) and ionized reflection from the inner disk (blue).}
    \label{fig:ngc3516}
\end{figure}

The Seyfert 1 galaxy NGC~3516 was observed as part of this program in November 2009, for a total on-source exposure time of 250\,ks.      This object is known to possess a very strong and complex warm absorber that has substantial variations on timescales of months \cite{turner11}.   Our preliminary work suggests that our 0.5--30\,keV {\it Suzaku} spectrum admits degenerate solutions, describable by either a multi-zone partial covering warm absorber or an absorber plus an additional disk reflection component.    This degeneracy remains when one breaks the observation into four equally-spaced segments (during which the source monotonically decreases in flux by a factor of almost three).   Comparison of these four data segments shows that, once we account for a time-invariant distant reflector, the source spectrum maintains a fixed shape as it fades (Fig.~\ref{fig:ngc3516} [left]) --- thus, unless it is takes the form of sharp-edged Compton-thick ``bricks" \cite{turner11}, the absorption must be essentially fixed during the 460\,ks of this observation.   We stress that the presence of strong disk reflection signatures {\it is} consistent with these data (Fig.~\ref{fig:ngc3516} [right]) --- however, due to the degeneracies with the complex absorber, we are not currently able to report any spin measurements for this object.   Our detailed analysis of these data will be described in a forthcoming publication.

\section{Discussion and conclusions}

\begin{figure}
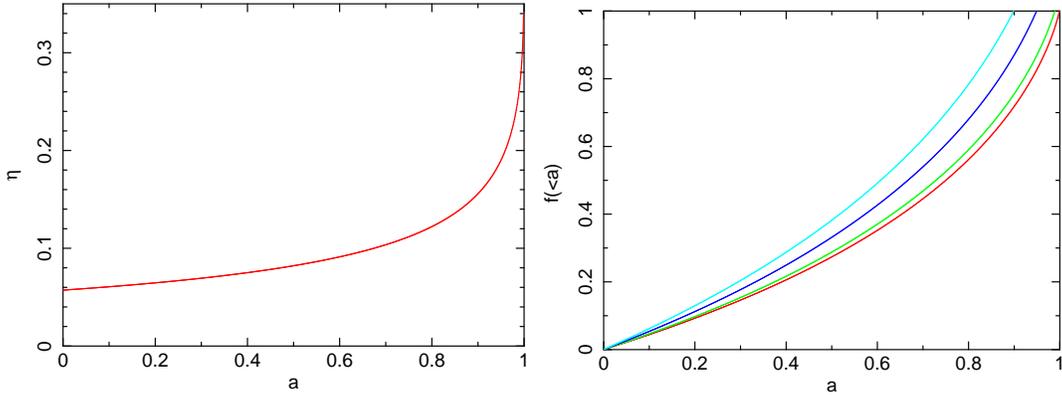

\hbox{
  \includegraphics[height=.32\textheight,angle=270]{efficiency.ps}
  \includegraphics[height=.32\textheight,angle=270]{selection_fig2.ps}
  }
  \caption{{\it Left panel : }Radiative-efficiency $\eta$ as a function of black hole spin for a standard accretion disk assuming a zero-torque inner boundary condition, and not accounting for photons that are captured by the black hole (a small effect).  {\it Right panel : }Illustration of spin bias in a flux limited sample induced by the spin dependence of the radiative efficiency.   Here, we assume that the intrinsic population of black holes has a uniform spin distribution for spins in the range $a=0\rightarrow a_{\rm max}$.   We then show the cumulative fraction of sources in a flux limited sample that possess a spin less than $a$.   Shown here are the cases $a_{\rm max}=0.998$ (red), $a_{\rm max}=0.99$ (green), $a_{\rm max}=0.95$ (blue), and $a_{\rm max}=0.9$ (cyan).   Based on discussion and analysis presented in \cite{brenneman11}.  }
  \label{fig:bias}
\end{figure}

Combining our Key Project results with those from other groups, the number and robustness of SMBH spin measurements has improved dramatically just over the past two years (see Table~2 of \cite{brenneman11}).  While still rudimentary at the current time, we are approaching a time when the spin-distribution function of SMBHs in the brightest/nearest AGN can be assessed observationally --- this would provide a new dimension in which to constrain models of SMBH growth and evolution \cite{volonteri05}.  

Of the eight AGN that currently have iron-line spin constraints, only Fairall~9 appears to be consistent with a SMBH that is spinning less that $a=0.5$.   However, before concluding that the local SMBH spin-distribution is weighted towards high-spins, we must ask whether there are spin-dependent selection effects that will bias the measured spin distribution.     Even in a simple picture, it is easy to see that a flux limited sample will be biased towards high spins.   For standard accretion models, the efficiency of black hole accretion increases as the spin of the black hole increases (Fig.~\ref{fig:bias} [left]). So, all else being equal, an accreting, rapidly spinning black hole will be more luminous than an accreting, slowing spinning black hole and hence will be overrepresented in flux-limited samples.  To illustrate that this is a significant effect, Fig.~\ref{fig:bias} [right] shows the cumulative fraction $f(<a)$ of objects in a flux limited sample has a spin less than $a$ when the intrinsic population has a uniform distribution of spins between 0 and $a_{\rm max}$.    For $a_{\rm max}=0.99$, we find that half of the sources in a flux limited sample have spin $a>0.73$, and only 28\% of the sources in the sample have $a<0.5$ (whereas 50\% of the parent population has $a<0.5$).  Of course, this particular effect can be considered the ``minimal" selection bias.   A more rigorous assessment of spin-biases must include spin-effects on the X-ray source structure as well as the ionization and density structure of the inner accretion disk.   It is tremendously exciting that these questions, so long the purview of the theorists, are now the subject of observational studies.

%


\begin{theacknowledgments}
CSR is grateful to the organizers of this conference for the opportunity to present these results.   We thank both the US and Japanese {\it Suzaku} teams for enabling this Key Project.   CSR acknowledges support from the NASA Suzaku Guest Observer Program under grants NNX09AV43G and NNX10AR31G.   RCR is supported by NASA through the Einstein Fellowship Program, grant number PF1-120087.
\end{theacknowledgments}



\bibliographystyle{aipproc}   


\begin{thebibliography}{0}
\expandafter\ifx\csname natexlab\endcsname\relax\def\natexlab#1{#1}\fi
\providecommand{\enquote}[1]{``#1''}
\expandafter\ifx\csname url\endcsname\relax
  \def\url#1{\texttt{#1}}\fi
\expandafter\ifx\csname urlprefix\endcsname\relax\def\urlprefix{URL }\fi
\providecommand{\eprint}[2][]{\url{#2}}

\end{thebibliography}


\begin{thebibliography}{9}

\bibitem{brenneman06}L.W.Brenneman, C.S.Reynolds, 2006, ApJ, 652, 1028
\bibitem{brenneman11}L.W.Brenneman et al., 2011, ApJ, 736, 103
\bibitem{cackett10}E.Cackett et al., 2010, ApJ, 720, 1325
\bibitem{dabrowski97}Y.Dabrowski et al., 1997, MNRAS, 288, L11
\bibitem{dauser10}T.Dauser, J.Wilms, C.S.Reynolds, L.W.Brenneman, 2010, MNRAS, 409, 1534
\bibitem{davis11}S.W.Davis, A.Laor, 2011, ApJ, 728, 98
\bibitem{emm11}D.Emmanoulopoulos et al., 2010, MNRAS, 415, 1895
\bibitem{fabian95}A.C.Fabian et al., 1995, MNRAS, 277, L11
\bibitem{krolik02}J.H.Krolik, J.F.Hawley, 2002, ApJ, 573, 754
\bibitem{lohfink11}A.M.Lohfink et al., 2011, in prep
\bibitem{miller09}J.M.Miller, C.S.Reynolds, A.C.Fabian, G.Miniutti, L.C.Gallo, 2009, ApJ, 697, 900
\bibitem{miniutti07}G.Miniutti et al., 2007, PASJ, 595, 315
\bibitem{netzer2003}H.Netzer et al., 2003, ApJ, 599, 933
\bibitem{orban11}G.Orban~de~Xivry et al., 2011, MNRAS, in press
\bibitem{patrick10}A.R.Patrick et al., 2011, MNRAS, 411, 2353
\bibitem{patrick11}A.R.Patrick et al., 2011, MNRAS, 416, 2725
\bibitem{reis11}R.C.Reis et al., 2011, ApJ, submitted
\bibitem{reynolds08}C.S.Reynolds, A.C.Fabian, 2008, ApJ, 675, 1048
\bibitem{ross05}R.R.Ross, A.C.Fabian, 2005, MNRAS, 358, 211
\bibitem{schmoll09}S.Schmoll et al., 2009, ApJ, 703, 217
\bibitem{sikora07}M.Sikora, L.Stawarz, J.P.Lasota, 2007, ApJ, 658, 815
\bibitem{soltan82}A.~Soltan, 1982, MNRAS, 200, 115
\bibitem{tanaka95}Y.Tanaka et al., 1995, Nature, 375, 659
\bibitem{turner11}T.J.Turner, L.Miller, S.B.Kraemer, J.N.Reeves, 2011, ApJ, 733, 48

\bibitem{volonteri05}M.Volonteri et al., 2005, ApJ, 620, 69










\end{thebibliography}



\end{document}